# A Regression Analysis Study Examining the Price of Drones

Burchan Aydin, Subroto Singha
*Department of Engineering & Technology*
*Texas A&M University-Commerce*

## Abstract

Objective of this study is to examine the impact of twelve specifications namely No. of Rotors, Max Flight Time(min.), Operating Range (m), Wingspan(mm), Weight(g), Payload Capacity(g), Max Speed(m/s), Max Flying Altitude(m), GPS Compatibility, Autonomous Flight, Collision Avoidance, Flight Battery (mAh) on the price of the drones. Using Stepwise Multiple Linear Regression analysis, our study shows that at $\alpha=0.05$; Max Flight Time (min.) (p-value=<0.001), Wingspan (mm) (p-value=<0.001), and Autonomous Flight capability (p-value=0.006) are the significant impact factors, while GPS Compatibility (p-value=0.057), and Collision Avoidance (p-value=0.077) are marginally significant factors affecting the price of drones.

## Keywords

## 1. Introduction

Unmanned Aerial Vehicles (UAV), commonly known as "Drones", are extensively applicable in wide range of areas such as remote sensing [1], aerial photography [2], precision agriculture [3], reforestations [4], firefighting [5], healthcare [6], surveillance [7], military [8] etc. According to a report from DroneDeploy, usage of UAVs increased by 239% in 2018, compared to the previous year [9]. The Federal Aviation Administration (F.A.A.) forecasts a total hobbyist fleet of 3.55 million drones by 2021 in a base scenario, and 4.47 million drones in a best scenario. In addition, F.A.A. forecasts 422,000 commercial drones by 2021 in a base scenario and 1,616000 commercial drones in a best-case scenario [10]. In fact, the market size in the USA was worth of 2640 million USD in 2018 and is forecasted to reach 16,200 million USD by 2025 [11]. Despite the huge current, and predicted market, suitable cost metrics do not exist, which is necessary to determine the price of drones [12]. Identical drone models are placed in very different price ranges in the market. When deciding what model of drone to purchase, there is no clarity on the price range to look for. The price of commercially available drones ranges from $30 to $300,000. It is difficult for recreational, commercial, and public safety users to determine which drone to select based on their needs. It is vague if the price of the drones is impacted by certain specifications or is it all decisions of the sellers without any basis. By taking average price for the same drone model from 3 vendors, this study examined the statistically significant and marginally significant price affecting specifications of drones. Twelve different specifications of a hundred drone models existing in the market were collected and analyzed by Multiple Regression Analysis. This paper is organized as follows. The next section illustrates the methodology used in this study. Section 3 demonstrates the results, section 4 is the discussion of the results, and section 5 is the conclusion.

## 2. Methodology

One hundred drone models' specifications were collected including but not limited to the most popular drone companies such as DJI, Tarot, Yuneec, Dragonfly etc. Each drone model's following specifications were regarded as independent variables: Number of rotors, maximum flight time (min.), operating range (m), wingspan (mm), weight (g), payload capacity(g), maximum speed (m/s), max flying altitude(m), GPS compatibility, autonomous flight, collision avoidance, and flight battery (mAh). Several specification were eliminated from analysis due to excessive amounts of missing data such as body material, rotor material, ready to fly capability, 'altitude hold' and 'position hold' flight modes ability, 'follow me', 'return to home', 'auto landing' capabilities, and camera specifications such as resolution, and stabilization.

Among twelve specifications GPS Compatibility, Autonomous Flight, and Collision Avoidance were the categorical variables namely 'Yes' & "No'. In addition, number of rotors was accounted as a categorical variable (tricopter, quadcopter, hexacopter, and octocopter, coded as 0,1,2,3 respectively). It should be noted that fixed wing aircraft is not included in this study. The dependent variable is the price of the drones. It is a continuous variable. The average of the prices of a brand-new drone model from three different vendors were taken to avoid any bias. Data was collected in MS. Excel, then Python's Pandas library and Minitab 2019 were used in conjunction for data cleansing [13]. In order to examine the specifications affecting the price of drones, a stepwise multiple linear regression analysis was conducted in Minitab, 2019. Figure-1 demonstrates the methodology utilized in this study.

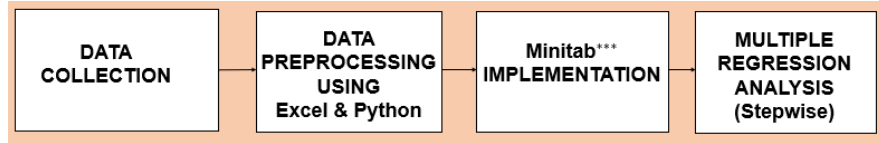

**Figure 1**: Methodology

The following section illustrates the results of the analysis.

## 3. Results

The descriptive statistics for the dependent variable are provided in Table-1 below. It can be observed that there is an enormous variability. However, the outliers were not removed from the analysis after detailed analysis of these two points. Nothing inaccurate was found regarding the data set for the outliers.

**Table 1**-Descriptive Statistics for the Price

| Variable | Mean | StDev | Median | Mode | N for Mode |
|---|---|---|---|---|---|
| Price ($) | 9199 | 3,169 | 1059 | 300 | 4 |

No violation of Normality assumption was detected as can be seen in Figure-2. However, two of the independent variables, weight, and payload capacity were found to be significantly correlated to each other. The weight variable was also correlated with other specifications; therefore, it was removed from further analysis.

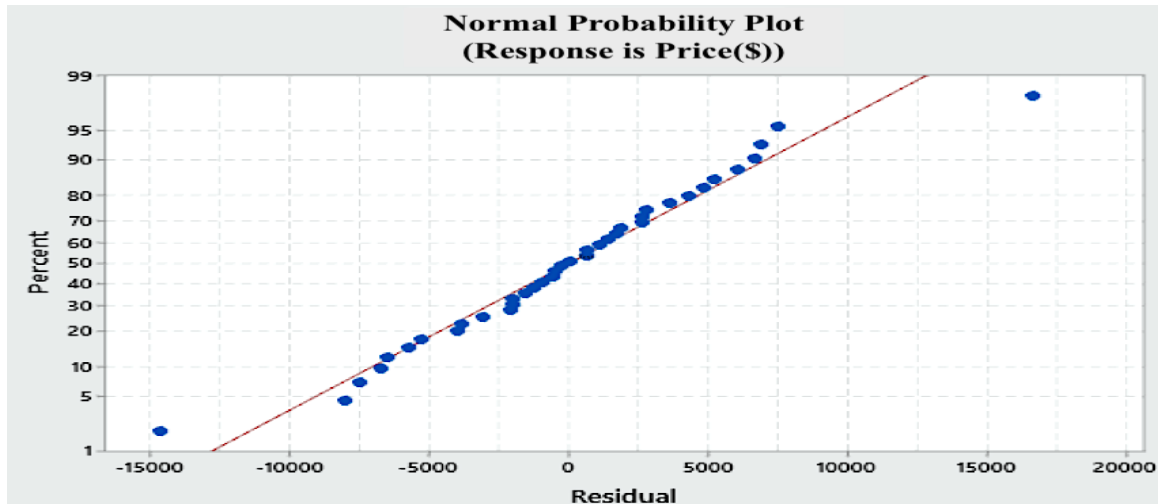

**Figure 2**: Normal Probability Plot

A multiple linear regression was calculated to predict the price of a drone based on twelve specification of drones as aforementioned. A significant regression equation was found ($F(5,32)=22.37$, $p <0.001$), with a $R^2$ of .7775. The final regression equation is shown below in Figure-3, and the ANOVA analysis results are shown in Figure-4.

$$y = Price; x_1 = Max\ Flight\ time(min); x_2 = Wingspan(mm); x_3 = GPS\ Compatible; x_4 = Automous\ Flight; x_5 = Collision\ Avoidance$$

$$y = -5205 + 354.7x_1 + 16.84x_2 - 12739x_3 + 7367x_4 - 4364x_5$$

**Figure 3:** Regression Model

| Analysis of Variance | | | | | |
|---|---|---|---|---|---|
| Source | DF | Adj SS | Adj MS | F-Value | P-Value |
| Regression | 5 | 3924201473 | 784840295 | 22.37 | 0 |
| Max Flight Time(min) | 1 | 760747416 | 760747416 | 21.68 | 0 |
| Wingspan(mm) | 1 | 1500447513 | 1500447513 | 42.77 | 0 |
| GPS Compatible | 1 | 137061748 | 137061748 | 3.91 | 0.057 |
| Autonomous Flight | 1 | 306602388 | 306602388 | 8.74 | 0.006 |
| Collision Avoidance | 1 | 117560062 | 117560062 | 3.35 | 0.077 |
| Error | 32 | 1122735313 | 35085479 | | |
| Total | 37 | 5046936786 | | | |

**Figure 4:** ANOVA Analysis Output

Figure-5 shows the p-values for the coefficients found in the model. Maximum flight time, wingspan, and autonomous flight were the significant factors at α=0.05, whereas GPS compatibility, and collision avoidance were the marginally significant factors at α=0.05. A legend with green color is used to show the significant factors and yellow color is used to show the marginally significant factors. Figure 5 also shows the variance inflation factors (VIF) for the independent variables. VIF values show no multi-collinearity in the final regression model, as all values are less than 7 [14].

| Coefficients | | | | | |
|---|---|---|---|---|---|
| Term | Coef | SE Coef | T-Value | P-Value | VIF |
| Constant | -5205 | 6000 | -0.87 | 0.392 | |
| Max Flight Time(min) | 354.7 | 76.2 | 4.66 | 0 | 1.23 |
| Wingspan(mm) | 16.84 | 2.58 | 6.54 | 0 | 1.29 |
| GPS Compatible | -12739 | 6445 | -1.98 | 0.057 | 1.15 |
| Autonomous Flight | 7367 | 2492 | 2.96 | 0.006 | 1.38 |
| Collision Avoidance | -4364 | 2384 | -1.83 | 0.077 | 1.33 |

**Figure 5:** Coefficients of the Regression Output

Using Durbin-Watson statistics, existence of any potential autocorrelation was also checked. The range of this statistic is 0 to 4. If the value is close to 2, then there is no autocorrelation [15]. Therefore, a Durbin-Watson statistic of 1.52877 indicates that no autocorrelation was detected.

## 4. Discussion

Although the aforementioned regression model might seem significant, the $R^2$ value is .7775. Thus, more than 20% of the variability of the drone prices are not explained by the specifications used in this analysis. Future studies might introduce more independent variables that could be affecting the drone prices. One of these factors might also be the country of origin for either the parts or the whole drone. The reason is that global trade regulations and costs, and the varying labor costs in different countries could be affecting the price of the drones, no matter what the specifications are. Also, prices differ based on companies' selling strategies. This study addressed this issue by taking average prices from three vendors.

As stated earlier, several variables were removed due to lack of data. Some companies do not share all specifications of their drones, which made it impossible to even get close to use data cleansing methods to fill the null values for those twelve specification categories. Some of these eliminated specifications were the material of the frame, and camera related specifications including but not limited to resolution, and gimbal stability and range of motion. It could be a future study to find the missing values and re-running the regression analysis with the new independent variables.

Moreover, from a statistical standpoint, different regression models could be selected such as polynomial regression to reflect more of the variability in the drone prices. Over twenty different specifications were initially determined, however for only twelve of these specifications enough data points were found by the research team.

It should be noted that collision avoidance decreased the price of the drones in the regression model, yet in reality this should not be favorable. Collision avoidance was only marginally significant together with the GPS compatibility; thus, they could be removed from the final model. However, both the collision avoidance and the GPS compatibility are crucial specifications of drones, thus, the authors decided to keep them in the final model even though they are contributing negatively to the model. Especially, collision avoidance is a very important specification for a drone, where only few sophisticated models such as DJI family possess this capability. In a future study, impact of collision avoidance on the price of the drone will be re-analyzed to understand why drone prices were lower in the data collected for this study.

On the other hand, the study provided several important findings to consider. Maximum flight time, wingspan, and autonomous flight specifications are the significant factors that affect the price of the drones. Drones are used for at least 40 different applications in commercial and public safety operations as listed in [16]. Each of these applications require different drone specifications. For instance, delivery drones need high payload capacity, maximum flight time, autonomous flight, and collision avoidance. Whereas, a drone to be used for real estate photography only needs a high-resolution camera. Considering these results, as long as the user does not need long flight time, a high wingspan, or autonomous flights, affordable drones could be purchased. The results of this study indicated that decision makers should consider their needs before purchasing a drone as the price variability is significantly high. This could create the opportunity of a better return on investment, no matter what the application it is purchased for.

## 5. Conclusion

This study examined the factors affecting the price of drones commercially available in the market. As with all technology, it is certain that there are several factors that could be affecting the price of drones, yet the scope of this study covered solely the impact of the specifications of drones. Maximum flight time, wingspan, and autonomous flight capability were found to the bottom-line specifications that affected the price of the commercially available drones. As long as an application do not require these three variables, a more affordable drone purchase could be made by decision makers.